\documentclass[twocolumn]{revtex4}
\usepackage{graphicx}
\usepackage{amssymb,amsfonts,amsmath}

\begin{document}

\title{Universal Robotic Gripper based on the Jamming of Granular Material}

\author{Eric Brown$^1$
Nicholas Rodenberg$^1$,
John Amend$^2$,
Annan Mozeika$^3$
Erik Steltz$^3$, 
Mitchell R. Zakin$^4$,
Hod Lipson$^2$,
Heinrich M. Jaeger$^1$
\\
$^1$ James Franck Institute, The University of Chicago, Chicago, IL 60637\\
$^2$ School of Mechanical and Aerospace Engineering, Cornell University, Ithaca, NY 14853\\
$^3$ iRobot G\&I Research, 8 Crosby Drive, Bedford, MA  01730,\\
$^4$ Defense Advanced Research Projects Agency, 3701 North Fairfax Drive, Arlington, VA 22203}

\begin{abstract} 

Gripping and holding of objects are key tasks for robotic manipulators. The development of universal grippers able to pick up unfamiliar objects of widely varying shape and surface properties remains, however, challenging.  Most current designs are based on the multi-fingered hand, but this approach introduces hardware and software complexities. These include large numbers of controllable joints, the need for force sensing if objects are to be handled securely without crushing them, and the computational overhead to decide how much stress each finger should apply and where. Here we demonstrate a completely different approach to a universal gripper. Individual fingers are replaced by a single mass of granular material that, when pressed onto a target object, flows around it and conforms to its shape.  Upon application of a vacuum the granular material contracts and hardens quickly to pinch and hold the object without requiring sensory feedback. We find that volume changes of less than 0.5\% suffice to grip objects reliably and hold them with forces exceeding many times their weight.  We show that the operating principle is the ability of granular materials to transition between an unjammed, deformable state and a jammed state with solid-like rigidity. We delineate three separate mechanisms, friction, suction and interlocking, that contribute to the gripping force. Using a simple model we relate each of them to the mechanical strength of the jammed state. This opens up new possibilities for the design of simple, yet highly adaptive systems that excel at fast gripping of complex objects.

\end{abstract}

\maketitle


Tasks that appear simple to humans, such as picking up objects of varying shapes, can be vexingly complicated for robots. Secure gripping not only requires contacting an object, but also preventing potential slip while the object is moved. This can be achieved either by friction from contact pressure or by exploiting geometric constraints, for example by placing fingers around protrusions or into the opening provided by the handle of a cup. For reliable robotic gripping, the standard design approach is based on a hand with two or more fingers \cite{PY91}-\cite{Bi00}, and typically involves a combination of visual feedback and force sensing at the fingertips. A large number of optimization schemes for finger placement as well as the use of compliant materials for adaptive grasping have been discussed \cite{Bi00}-\cite{CK06}.  Given the evolutionary success of the multi-fingered hand in animals, this approach clearly has many advantages.  However, it requires a central processor or ÒbrainÓ for a multitude of decisions, many of which have to be made before the hand even touches the object, for example about how wide to spread the fingers apart.  Therefore, a multi-fingered gripper not only is a complex system to build and control, but when confronted with unfamiliar objects it may require learning the shape and stiffness of the object. 

The focus of this work is on the problem of gripping, not manipulation, and seeks to offload system complexities such as tactile sensing and computer vision onto novel mechanical design.  This approach replaces individual fingers by a material or interface that upon contact molds itself around the object.  Such a gripper is universal in the sense that it conforms to arbitrary shapes and is passive in that all shape adaptation is performed autonomously by the contacting material and without sensory feedback.  This reduces the number of elements to be controlled and therefore can have advantages in terms of reliability, cost and gripping speed. So far, however, passive universal grippers have remained largely unexplored. An early snake-like gripper by Hirose \cite{HU78} employed a system of joints and pulleys with a single actuator.  A few designs have envisioned systems where moveable jaws with highly compliant surfaces contact the object from two or more sides, partially enveloping and thus securing it. For example, Choi and Koc recently presented a gripper whose jaws were outfitted with inflatable rubber pockets \cite{CK06}. Earlier, Schmidt \cite{Sc78} and Perovskii \cite{Pe80} introduced the idea of attaching elastic bags loosely filled with granular material, such as small pellets or spheres, to the gripper jaws. A similar idea was also put forward by Rienm\"uller and Weissmantel \cite{RW88}. These bags conform to the shape of any object they press against and, by simply evacuating the gas inside, can be turned into rigid molds for lifting the object. However, the mechanism for this transformation was not understood and no data about gripping performance were presented.  As a result, these early approaches to passive universal grippers never gained traction.

\begin{figure*}                                                
\centerline{\includegraphics[width=5.5in]{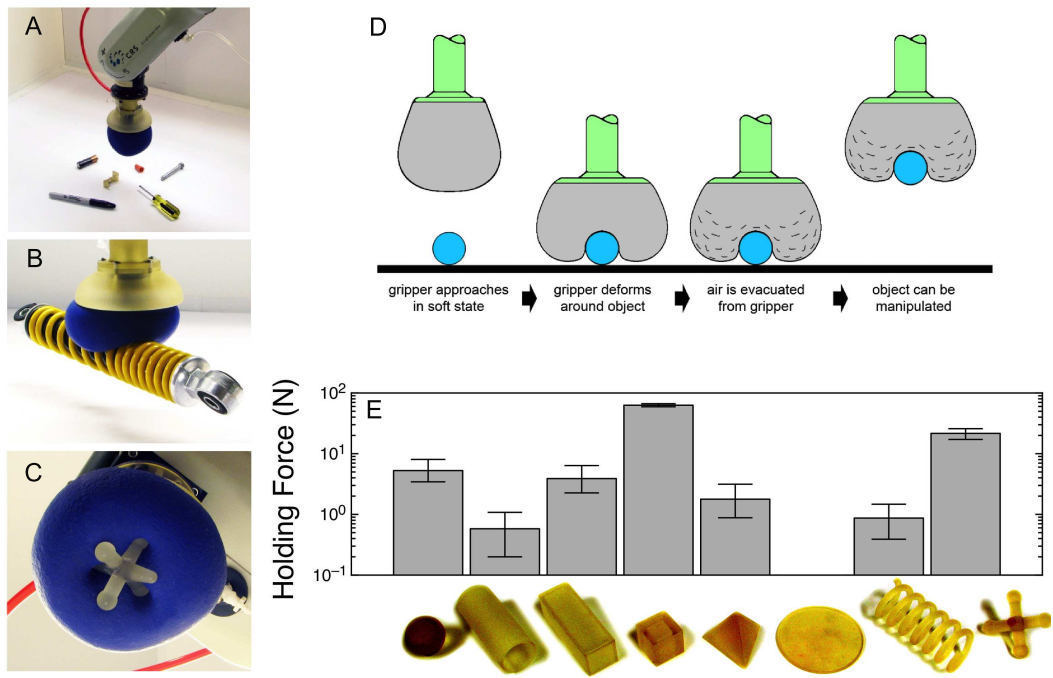}}
\caption{Jamming-based grippers for picking up a wide range of objects without the need for active feedback. (A) Attached to a fixed-base robot arm. (B) Picking up a shock absorber coil. (C) View from the underside. (D) Schematic of operation. (E) Holding force $F_h$ for several 3D-printed test shapes (the diameter of the sphere shown on the very left, $2R = 25.4$ mm, can be used for size comparison). The thin disk could not be picked up at all. }
\label{fig:Fig1}                                        
\end{figure*}

Here we revisit the idea of using granular material for a universal gripper and show that the gripping process is controlled by a reversible jamming transition \cite{MSLB07}-\cite{CWBC98}. While the concept of jamming has been used to explain the onset of rigidity in a wide range of amorphous systems from molecular glasses to macroscopic granular materials \cite{OSLN03, LN98, TPCSW01}, the benefits of jamming for the assembly of materials with tunable behavior are just beginning to be explored. The unique properties of a jamming gripper derive from the fact that loose grains in a bag sit at the threshold between flowing and rigid states \cite{JNB96}.  This enables the gripper to deform around the target in the unjammed, malleable configuration, then harden when jamming is initiated. In the vicinity of the jamming transition very small modifications of the packing density can drive dramatic changes in the mechanical response \cite{OSLN03,LN98}. Thus, increasing the particle confinement slightly, e.g., by applying a vacuum, enables the gripper to gain remarkable rigidity while almost completely retaining its shape around the target.

We focus on the simplest form of a gripper, a single non-porous elastic bag filled with granular matter (Fig.~1). This system approximates the limit of a robotic hand with infinitely many degrees of freedom, which are actuated passively by contact with the surface of the object to be gripped and are locked in place by a single active element, a pump that evacuates the bag. Figure 1 demonstrates that a wide range of different types of objects are easily handled in pick-and-place operations using a fixed-base robotic arm, without the need to reconfigure the gripper or even position it precisely, as long as it can cover a fraction of a target object's surface. This includes switching between objects of different shapes, items difficult to pick up with conventional universal grippers or fragile targets like raw eggs, as well as simple manipulation tasks, such as pouring water from a glass or drawing with a pen (see supplementary videos online).  The same type of gripper can also pick up multiple objects simultaneously and deposit them without changing their relative position or orientation.  For all of the items depicted in Fig.~1, we find that holding forces  can be achieved that exceed significantly the weight of objects of that size.   We find that this strength is due to three mechanisms, all controlled by jamming, that can contribute to the gripping process: geometric constraints from interlocking between gripper and object surfaces, static friction from normal stresses at contact, and an additional suction effect, if the gripper membrane can seal off a portion of the object's surface.

\section{Results and Discussion}

To evaluate gripping performance we performed pick-and-place operations in which objects were gripped, lifted and moved (Fig. 1D).  In addition, the holding forces required to pull out the objects were measured (Fig.~1E).  These tests were done with a fixed-base robotic arm to which a gripper bag of radius $L = 4.3$ cm was attached, containing ground coffee as the granular material (Fig.~1A-C). The bag was filled almost completely but not stretched out so the grains remained loosely packed and the gripper was malleable when no vacuum was applied. By establishing a differential jamming pressure $P_{jam}$ across the bag's latex rubber membrane (0.3 mm in thickness) the packing could be jammed. Employing a Venturi aspirator, compressed air was used to generate pressures $P_{jam}$ around 75 kPa, i.e., the bag was evacuated down to $\approx 1/4$ atm -- a level easily reachable with a small vacuum pump. For a wide range of objects, including those shown in Figs.~1A and E but also small flashlight light bulbs, M\&Ms$^{\textregistered}$, LEDs, bottle caps, plastic tubing, foam ear plugs, and a variety of hardware items and office supplies, the pick-up success rate in 10 trials each was 100\%. The magnitude of the holding force, however, was clearly influenced by the objects' shape (Fig.~1E). The only objects that could not be gripped were those in which the gripper membrane could not reach sufficiently around the sides, e.g., for hemispheres larger than about half the size of the gripper or for thin disks lying flat, or for very soft objects like cotton balls. 

\begin{figure}                                                
\centerline{\includegraphics[width=2.5in]{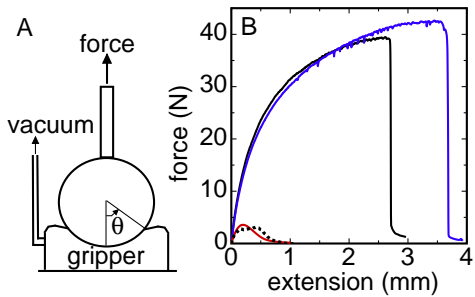}}
\caption{(A): Sketch of setup to measure the holding force. (B): Measured force as the sphere is pulled up vertically out of the gripper. Data shown are for fixed contact angle $\theta = \pi/2$, sphere diameter $R = 12.5$ mm, and confining pressure $P_{jam} = 80$ kPa but different sphere surfaces: solid and dry (black), solid and moistened (blue), solid and powdered with cornstarch (red), porous (dotted). Spheres that form a airtight seal with the gripper membrane are held with a force about 10 times that of porous spheres or those with powdered surfaces that do not seal as well.}
\label{fig:holdingforce}                                        
\end{figure}

In the following we focus on spheres as test objects to isolate contributions from individual gripping mechanisms and perform quantitative comparisons with model predictions.  The gripper used for these holding force measurements was stationary and consisted of a rubber bag (0.3 mm in thickness) with average $L \approx 4$ cm, filled with smooth soda-lime glass spheres 100 $\mu$m in diameter to about 80\% of the bag volume. The experiment used an inverted configuration, in which the target object, an acrylic sphere with radius $R$, was attached to an Instron 5869 materials tester and pressed into the gripper bag which itself was fixed to a flat surface.  A differential jamming pressure $P_{jam}$ was then applied.  The holding strength was measured by pulling the sphere out of the gripper and recording the tensile force as a function of vertical extension.  A diagram of this setup and typical force-extension curves for are shown in Fig.~2.  Additional measurements showing that the gripper also resists lateral forces as well as torques (required for force closure \cite{PF95, CK06, MLS94, Yo99}) are presented in the supplementary information. The gripping performance was investigated for different $P_{jam}$, $R$, and surface properties, although for brevity we focus here on data taken with $P_{jam} = 80$ kPa and $R= 19$ mm.

Focusing on the maximum force that can be sustained prior to failure, one of the features seen in Fig.~2 is the enhancement of the holding strength when the interface between the sphere and the rubber seals tightly for wet or dry smooth surfaces.  This seal is key for the suction effect between the gripper and the sphere. When it cannot be established, shown for the cases of a porous sphere or a surface roughened by a coating with $\approx 20$ $\mu$m diameter powder particles, the holding force drops significantly. 

\begin{figure*}                                                
\centerline{\includegraphics[width=6.25in]{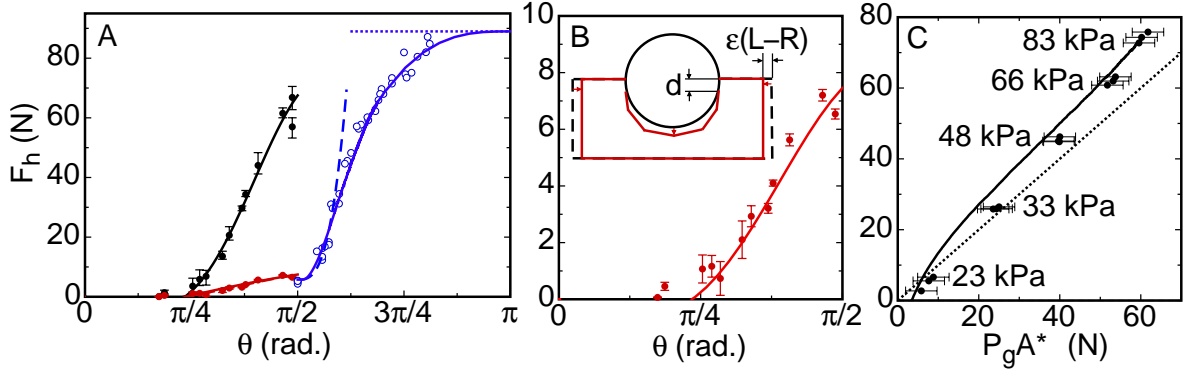}}
\caption{Gripper holding force $F_h$. (A)  $F_h$ as a function of contact angle $\theta$.  Data are for a porous sphere gripped by friction for contact angles between $\pi/4$ and $\pi/2$ (red) and by geometrical interlocking for contact angles above $\pi/2$ (blue), and for a solid sphere held by both friction and suction for contact angles between $\pi/4$ and $\pi/2$ (black).  Error bars indicate range of forces obtained from 5 repeated measurements.    The uncertainty in $\theta$ is 0.05 rad.  Lines give predictions as discussed in the text. (B) Zoomed-in version of the data in (A) for friction.  Inset: Sketch of the contraction that occurs when a gripper bag of diameter $L$ jams around a sphere, producing an O-ring-like pinching region of width $d$ (not to scale). (C) Holding force $F_h$ for solid spheres as a function of vacuum pressure $P_g$ in the gap below the sphere, where the gap extends over a horizontal cross-sectional area $A^*$. (solid line): Build-up of $P_g$ during a single run at $P_{jam} = 80$ kPa in which the sphere was pulled slowly until $F_h$ was reached. (Solid symbols): Maximum holding force $F_h$ in relation to gap pressure $P_g$ for different confining pressures $P_{jam}$ listed on the plot.  (dotted line):  $F_s = P_g A^*$.   All data are for $R = 19$ mm spheres. }
\label{fig:threemechanisms}                                        
\end{figure*}

The degree to which the sphere is enveloped by the gripper is given by the contact angle $\theta$ (Fig.~2). Plotting the peak holding force, $F_h$, as a function of $\theta$, allows us to identify different gripping regimes (Fig.~3A, B). Below a minimum angle $\theta \approx \pi/4$ the gripping strength vanishes except for a small contribution from residual membrane stickiness. Above $\pi/4$ there is a rapid increase in $F_h$ with contact angle (red data points). As already seen in Fig.~2, the holding force is enhanced considerably if the sphere surface is smooth to allow for the suction mechanism to operate (black data points). This enhancement occurs with the same onset threshold as for the case without suction. Once the sphere is more than half enveloped and $\theta > \pi/2$, a new regime is entered in which geometric interlocking leads to significant additional holding strength (blue data points).

As the bag is evacuated, the differential pressure $P_{jam}$ across the membrane leads to a volume contraction of the particle packing, as indicated by the sketch in Fig.~3B. This contraction has two consequences:  it tightens the contact between the bag and the gripped object, and at the same time it jams and hardens the granular material inside the bag.  Because the packing density of granular material assembled under gravity is inherently at the threshold of jamming (a pile of grains can sustain a finite angle of repose) even a small applied confining stress $P_{jam}$ can frustrate the ability of grains to slip past one another and drive the packing deep into the jammed state.

For a quantitative modeling of the gripping action, we treat the gripper as an elastic medium and its volume change as producing stresses analogous to differential thermal contraction in a ball-and-socket joint.  The contraction of the gripper  is expected to pinch the surface of a hard sphere horizontally near the membrane-sphere contact line  as sketched in Fig.~3B.  This pinching applies a stress of magnitude $\sigma^*$  along a thin band of width $\delta\theta$ centered at $\theta$.  We assume that $\delta\theta$ is small and does not vary much with $\theta$.  Thus, the pinched region acts like an O-ring of width $d = R \delta\theta$ and diameter $2\pi R\sin\theta$, pressing against the sphere across an area $A_0 = 2\pi Rd\sin\theta$. The resulting normal force on the sphere, $F_N = \sigma^* A_0\sin\theta = 2\pi Rd\sigma^* \sin^2\theta$, gives rise to a tangential frictional force of magnitude $\mu F_N$ where $\mu$ is the static coefficient of friction at the membrane/sphere interface. Balancing the vertical components of these forces then gives the maximum vertical force that can be applied before the interface slips, $F_f = F_N (\mu\sin\theta - \cos\theta)$.  When the contact angle is less than a critical angle $\theta_c = \arctan(1/\mu)$ then $F_f = 0$ and the interface will slip regardless of the amount of applied normal force.  For rubber $\mu \approx 1$ and the contact angle must be greater than about $\pi/4$ for the frictional mechanism to work. For a porous sphere for which a vacuum seal cannot form the holding force thus is

\begin{equation}
F_h=F_f = 2\pi Rd\sigma^* (\mu\sin\theta-\cos\theta)\sin^2\theta	
\end{equation}

For a smooth solid sphere the effective O-ring can form an airtight seal, which can hold a vacuum in a gap in the region $\pm\theta$ inside the contact line.  To show this, we measure the gap pressure $P_g$ inside this region directly by using a sphere with a hole drilled through it and a vacuum gauge attached at the other end.  No pressure drop is detected when the gripper contracts around the sphere.  This demonstrates that the jamming serves the purpose of pinching the membrane to form a seal, but does not by itself generate a vacuum.  However, as the sphere is lifted, the jammed material inside the bag deforms, the gap starts to open, and $P_g$ builds up.  The resulting vertical suction force is $F_s = P_g A^*$ where $A^*$ is the horizontal cross-sectional area enclosed by the contact line. For a sphere $A^* = \pi R^2 \sin^2\theta$. Figure 3C demonstrates the suction effect in two ways:  First, by establishing that an interface pressure $P_g$ is built up as soon as the sphere is pulled on, and second, by  showing that the holding force $F_h \propto P_g$ as expected for suction.  The dotted line corresponds to $F_s = P_g A^*$, and the excess holding force above this line thus specifies the additional contribution from friction (about 20\%).

The pressure on the pinched O-ring will keep the vacuum seal in place as long as the frictional stress exceeds the gap pressure $P_g$.  Thus the maximum gap pressure before the seal fails is $P_g = F_f/A_0 = \sigma^*\sin\theta(\mu\sin\theta - \cos\theta)$.  This leads to a common onset threshold $\theta_c$ for gripping by either friction or suction, as borne out by the data in Fig.~3.  As a result, $F_s = P_g A^* = \pi R^2 \sigma^*  (\mu\sin\theta - \cos\theta) \sin^3\theta$ and the total holding force combining friction and suction is 

\begin{equation}
F_h  =  F_s + F_f  = \pi R^2 \sigma^*  (\mu\sin\theta - \cos\theta) \sin^3\theta \left(1 + \frac{2d}{R\sin\theta}\right).		
\end{equation}

\noindent Since $A^*/A_0 > 1$, the frictional term, i.e., the second term in the parentheses, typically makes only a small contribution to $F_h$ when a seal is formed.

From simultaneous fits of Eqs.~1 and 2  to the data for the porous and solid spheres, respectively, we find $\mu = 1.04 \pm 0.06$, $\sigma^* =50\pm4$ kPa, and $d = 1.07\pm0.07$ mm (these fits extend over the range $\pi/4 < \theta < \pi/2$  in Fig.~3).   The fit value for $\mu$ is consistent with the independently measured coefficient of friction $\mu=1.10\pm0.03$ (see Methods).  This, along with the fact that $d \ll R$, confirms the assumption that the pinching stress occurs in a thin region near the contact edge in a geometry resembling an O-ring.  A simple geometric model for the width of the pinched region on a sphere gives $d = [2\epsilon R(L-R)]^{1/2}$ in the limit $\epsilon \ll 1$ and results in $d=1.1$ mm, consistent with the fit value.  For typical granular materials, confining pressures $P_{jam}$ approaching one atmosphere lead to strains $\epsilon$ around one percent. Thus, the enhancement of the holding force due to suction is generally expected to be of order $\epsilon^{-1/2} \sim 10$ for spheres.  For other target shapes, the seal thickness $d$ will likely depend on the local curvature of the surface it is pressed against, with flatter surfaces allowing for  larger values of $d$.

The contraction stress $\sigma^*$ can be related to the strength of the jammed state. This is done by measuring the compressive strength of the jammed material with a triaxial compression test.  A stress-strain curve $\sigma(\epsilon)$ from such a test with a confining pressure $P_{jam} = 80$ kPa is shown in Fig.~4A.  To determine which point on the curve is relevant in the gripping experiments, a volumetric strain $\delta V/V$ is measured in the triaxial test cell as the confining pressure is applied.  For $P_{jam} = 80$ kPa, $\delta V/V = -0.004$  as shown in Fig.~4B.  Note that jamming is a reversible transition and that a similarly minute $\delta V/V$ suffices to drive the packing back into an unjammed configuration.  In fact, simply releasing the vacuum (Fig.~4B), even without any stirring or jostling, produces significant dilation of the packing and recovery of an easily malleable state.   Evaluating the compressive stress in Fig.~4A at a linear strain $\epsilon = (1/3)|\delta V/V|=0.0013$  gives 50 kPa.  This is in excellent agreement with the fit value for $\sigma^*$, and thus equates the compressive stress pinching the gripped target with the strength of the jammed material at the strain induced by $P_{jam}$.

\begin{figure}                                                
\centerline{\includegraphics[width=3.4in]{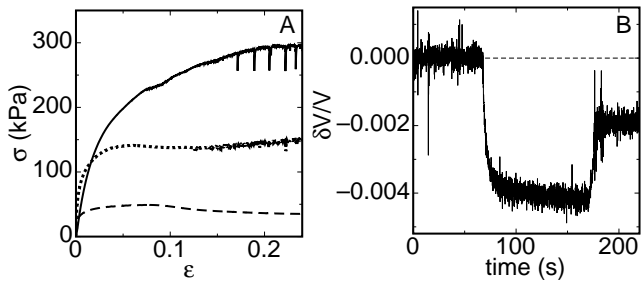}}
\caption{(A) Stress-strain curves $\sigma(\epsilon)$ from triaxial compression (dotted line), triaxial extension (dashed line), and 3-point bending (solid line) tests on 100 $\mu$m glass spheres in a latex membrane at $P_{jam} = 80$ kPa. (B) Volumetric strain $\delta V/V$ when a confining stress $P_{jam} = 80$ kPa is switched on at a time of 65 s and off at 170 s to jam and unjam the gripper. }
\label{fig:stressstrain}                                        
\end{figure}

If the contact angle $\theta > \pi/2$, gripper and gripped object have geometrically interlocked (satisfying form closure, see Refs.~\cite{CK06, MLS94, Yo99}). A gripper with an elastic membrane might conform to protruding parts of objects (as in Fig.~1C) to produce such interlocking, but the stiffness of the membrane usually prevents wrapping around convex objects.  To investigate the mechanism for interlocking quantitatively and in a simple geometry,  we therefore manually molded the jammed gripper around the porous sphere.  Then, to break the interlocking effect, the jammed material must both bend out of the way and stretch azimuthally to open enough to let the sphere through. Thus, we expect the holding force in this regime to depend on the resistance to a combination of bending and stretching.  Stress-strain curves measured from a 3-point bending test and a triaxial test for extension (stretching) of the granular material are shown in Fig.~4A.  While these curves differ in some details, they are both characterized by two key features:  a linear regime $\sigma=\epsilon E$ in the limit of small strains, where $E$ is the modulus, and at large strains a plateau around a level $\sigma_f$, the maximum stress the jammed material can sustain.  To understand the interlocking effect we first consider these two limits.

Where there is minimal interlocking, i.e.~$\theta - \pi/2 \ll 1$, the strain required to open up the gripper to allow the sphere to escape is small.  The minimum contribution from interlocking, $F_i$, to the holding force is the amount required to bend the ring wrapped around the sphere to vertical so the sphere can slip out.  In the small-$\epsilon$ limit, $\epsilon \approx 1-\sin\theta$ and  $F_i \approx (\pi/2)ER^2(t/l)^3(\theta-\pi/2)^3$, where $t$ is the thickness of the gripper section wrapped around the sphere and $l$ is the bending arm length.  Alternatively, to stretch open the neck of the region wrapped around the sphere so it can slip through requires a  force $F_i \approx (ERt/6)(\theta-\pi/2)^3$.   Because the location of the bend is not predetermined and the thickness is typically non-uniform, these are predictions for the scaling but can only provide a rough estimate for the magnitude.  Since $t$ and $l$ are typically comparable, we take $l \sim t \approx 5$ mm. This simplification gives the same scaling for both bending and stretching.  Since the bending resistance is seen to be considerably larger than the resistance to stretching for $\epsilon>0.003$ (Fig.~4A), corresponding to $\theta > 0.53\pi$, bending is expected to dominate the interlocking mechanism at larger $\theta$.  The stress-strain curve for bending in Fig.~4A is seen to be approximately linear for  $\epsilon<0.02$; by fitting we extract an effective bending modulus $E \approx 7.4$ MPa, and thus $(\pi/2)ER^2 = 4.2$ kN.  Fitting $F_h -F_f \propto  (\theta-\pi/2)^3$ to the data for $0.53\pi < \theta < 0.57\pi$ in this linear bending-dominated region we obtain the dashed line in Fig.~3A and a prefactor $(\pi/2)ER^2 = 1.6\pm0.3$ kN.  The fact that these two values for the force scale are of the same order of magnitude supports the notion that  the initial upturn in holding force for $\theta > \pi/2$ can indeed be attributed to the bending resistance of the jammed material.

In the opposite limit, for a high level of interlocking at large contact angles $\theta \gg \pi/2$, large strains will be required to pry open the bag. In this limit, the plateau of $\sigma(\epsilon)$ to $\sigma_f$ will cause $F_i$ to saturate for both bending and stretching.  The maximum force is then $\sigma_f$ times a bending area factor, which gives  $F_i \sim (2\pi R t^2/l) \sigma_f$.  Taking $\sigma_f = 0.29$ MPa from the stress-strain curve for bending (Fig.~4A) and $t = l$  leads to $F_i \sim 2\pi Rt \sigma_f \approx 170$ N, about twice the upper limit found in Fig.~3A (dotted blue line), again indicating that the scale of the maximum holding force due to interlocking is set by the maximum stress the jammed material can sustain under bending.

To capture the cross-over between these two limits, we use the full stress-strain curve $\sigma(\epsilon)$.  Integrating the stress over the bending area gives $F_i = \int_{\pi/2}^{\theta} (2\pi Rt^2/l) \sigma(\epsilon) \sin\theta' d\theta'$.   The stress can be evaluated using the small-$\epsilon$ limit $\epsilon \approx 1-\sin\theta$ because that is the only regime where where the stress-strain curve is still evolving.\footnote{In our gripper the moulding of the bag around the test sphere resulted in thinning near the opening, such that  $t$ decreased roughly linearly with $\theta$.  The effect of such thinning is that $F_i$ levels off at a somewhat lower value and at $\theta<\pi$.  To model the thinning, we can take $t\approx 2R(\pi-\theta)/\pi$.}  The resulting crossover is shown as the solid blue line in Fig.~3A, scaled by a factor of 0.23 to fit the data.   

During operation a grip may experience off-axis forces and torques, in addition to lifting forces discussed so far.  We show in the supplementary material holding forces measured for off-axis forces and torques.  We find that the friction mechanism is operative at about the same magnitude for resisting forces in all directions and torques applied at the surface.  Suction may be operative in some cases but this is dependent on the target geometry and force direction.

The above results demonstrate that the holding force for all three gripping mechanisms is directly related to the strength of the granular material in its jammed state:  contributions to $F_h$ from friction and suction are proportional to the pinching stress $\sigma^*$ that builds up as the contracting material compresses against the object to be gripped; contributions from geometric interlocking can involve the full stress-strain curve, depending on the extent of interlocking.  Since its rigidity is determined  by how deep the material is driven into the jammed state by the vacuum-induced volume contraction, the key control parameter for the gripping strength is the confining pressure $P_{jam}$.  In particular,  the confining pressure sets the overall scale for the stresses \cite{LW69} obtained from triaxial compression, 3-point bending, and stretching tests of the granular material as seen in Fig.~4A, so $\sigma^*$ and $\sigma_f$ are both the same order of magnitude as $P_{jam}$. Furthermore, the holding forces are approximately proportional to $P_{jam}$ (Fig.~3C).  While properties of the particles inside the bag such as shape and surface roughness can have a secondary contribution to the stress-strain curves (28) and thus the holding forces, we expect for all three mechanisms the maximum holding force should scale as $F_h \sim P_{jam} R^2$.


This scaling can be used to estimate the sizes of objects that can be lifted. Since the weight of a gripped object scales with volume, but the holding forces scale with area, we predict that the gripper can pick up objects up to a size of $R_{max}\sim  P_{jam}/(\rho g)$. For a typical metal ($\rho\approx 10^4$ kg/m$^3$) and $P_{jam} \approx 100$ kPa,  this gives an upper limit $R_{max} \sim 1$ m ($\sim 10^4$ kg) with either suction or interlocking.  For such big grippers, the weight of the granular material itself might become an issue but can be reduced by using hollow particles. Indeed, meter-size panels of vacuum-packed hollow spheres show remarkable stiffness and have been proposed as structural elements in architectural projects \cite{HH08, SLHS07, KKB08}.  While suction is not operative for all objects, interlocking is expected to be prevalent in a multi-bag, jaw-type gripper \cite{Sc78, Pe80, RW88}.  Still, even without suction or interlocking, friction alone makes it possible to grip and lift solid metal objects up to $R_{max} \sim 10$ cm.  Thus, friction provides more than enough force to pick up any of the objects shown in Fig.~1.

The above analysis was applied to spheres as test objects, but it allows us to draw some general conclusions.  For an arbitrarily shaped object, $\theta$ can be reinterpreted as the angle of a surface normal vector of the object where the pinching occurs. We can then re-write Eq.~2 in the form $F_h = \sigma^*\sin\theta (\mu\sin\theta - \cos\theta) [A^* + A_0]$, where it depends only on $\theta$, the pinching area $A_0$, and the horizontal cross-sectional area $A^*$ inside the pinching perimeter if a seal is formed. Both friction and suction require that the local slope at the contact line be steeper than $\theta_c = \arctan(1/\mu)$.  

With this model we can now explain the variation in holding forces measured in Fig.~1E.  The 3D-printed plastic material in this test is not smooth enough for the gripper to achieve an airtight seal.  Thus, the sphere is gripped by friction only and $F_h$ is in the range of what we see in Figs.~2B and 3A for porous spheres. The cylinder has a lower $F_h$ compared to a sphere of the same cross-section because it displaces a larger gripper volume which therefore does not reach down as far on the sides, resulting in a smaller vertical component of contact area $A_0\sin\theta$.  Despite its sharp edges, the cube is held with a large force in the range of what is observed for suction with smooth spheres.  The flat vertical faces allow for a large contact area from pinching comparable to the area that could be covered by suction, so the frictional effect is about as large as suction.  Compared to the cube, the vertical contact area of the cuboid is reduced, just as it is in the comparison between sphere and cylinder.  The tetrahedron presents a contact angle $\pi/3$ to the gripper, which explains the slightly reduced $F_h$ compared to the sphere.  The flat disk cannot be lifted since the gripper cannot get around the sides; thus the contact angle effectively is zero. The helical spring is similar to the cylinder in shape, and a similar lifting force is found.  The jack displays a larger force than can be expected from friction alone, indicating some amount of interlocking, as seen in Fig.~1C. 

Another aspect concerns the hardness of the object being gripped.  So far, we assumed the target was relatively hard so the stress response was solely determined by the gripper hardness.  However, for softer targets, the combination of the target and gripper must be considered in series.  A soft target will be strained as the gripper contracts, and the pinching pressure at the interface cannot exceed the strain of the gripper under vacuum times the target modulus. Thus, soft targets will experience less holding force.  Nevertheless, since friction is more than sufficient to lift hard objects on the cm scale (by a factor of about 30 for a density of 1g/mL), it should also hold soft targets with a modulus as small as order 1 MPa (about $1/30$ of the effective $E$ for compression in Fig.~4A). Indeed, foam earplugs were gripped readily by the set-up shown in Fig.~1, but not surprisingly one test object we failed to pick up was a cotton ball.

Neither the bag geometry nor details of the granular material seem to influence $F_h$ strongly, as long as they do not interfere with the degree to which the membrane can conform to an object's surface. In this regard, small grain size will be advantageous. However, very fine powders do not flow well and tend to stick. Furthermore, the permeability of a powder scales with the square of the grain diameter \cite{Mo05,PDB95}; thus, decreasing that diameter will increase the time required to reach a strongly jammed state. The membrane itself has to be sufficiently flexible and impermeable to allow for $P_{jam} > 0$.  For friction or suction to work at small contact angles a coefficient of friction $\mu\approx 1$ and some membrane elasticity are desirable, as in a rubbery material, but here we do not focus on optimizing the membrane (see Ref. 14 for a discussion of wear resistance of inflatable rubber pockets for robotic grippers). The gripping capabilities are therefore expected to be quite robust.

\section{Conclusions}

Our results demonstrate how minute changes in the packing density ($|\delta V/V| < 0.5\%$) associated with a vacuum-induced jamming/unjamming transition enable a universal granular gripper to adapt its shape to a wide range of different objects and pick them up reliably.  Without the need for active feedback, this gripper achieves its versatility and remarkable holding strength through a combination of friction, suction, and geometrical interlocking mechanisms. Only a fraction of an object's surface has to be gripped to hold it securely. Applied to spheres as test objects the simple model we introduced captures quantitatively the holding force for all three mechanisms.  Specifically, the model relates the gripping performance to the jamming pressure $P_{jam}$ and the stress-strain relationship of the granular material, and it predicts how the holding force scales  with object size, surface roughness (to the extent that an airtight seal can form), and surface normal angle at the gripper-object interface.

A universal gripper based on jamming may have a variety of applications where some of the high adaptability of a human hand is needed but not available or where feedback is difficult to obtain or expensive. Examples include situations where very different objects need to be gripped reliably and in rapid succession. A granular system can move with ease from gripping steel springs to raw eggs, and it can pick up and place multiple objects without changing their relative orientation.  Its airtight construction also provides the potential for use in wet or volatile environments. Another situation where such a gripper has a significant advantage over traditional designs is when minimal initial information is available, for example when the detailed shape or material properties of the target object are not known a priori, or when precise positioning is not feasible.  Since the gripper material adapts and conforms autonomously to the surface of the target object, a jamming-based system can be expected to perform particularly well for complex target shapes.

\section{Materials}
For pick-and-place performance evaluation we used a CRS A465 robotic arm, which includes high-pressure air lines, controlled by an imbedded solenoid valve. Ground coffee was chosen as the grain material for these tests because of its performance in jamming hardness tests.  The relatively low density of ground coffee is also advantageous, as it can be used to fill relatively large grippers without weighing them down and straining the membrane. The items shown in Fig. 1E were fabricated from photocurable plastic using an Objet 3D printer.  For the compressive stress-strain curves and the volumetric strain measurements (Fig.~4) a triaxial test cell (Durham Geo S-510A) was used and the granular material was contained in a 0.6 mm thick cylindrical rubber sleeve (51mm inner diameter). For bending tests a cylindrical sample 0.3 mm thick with 35.6 mm inner diameter was used in a standard 3-point test fixture. The volumetric strain $\delta V/V$ was obtained by measuring water displacement in the volume surrounding the rubber sleeve while applying vacuum to the interior of the sleeve.  The coefficient of friction $\mu$ between the acrylic and rubber membrane was obtained by fits to Eqs. 1 and 2 and also measured independently.  This was done by an inclined-plane test with four acrylic spheres taped together to prevent rolling on a rubber surface with an applied load of $200$ kPa, resulting in $\mu = 1.10 \pm 0.03$.  The fact that this value is slightly larger than unity is likely caused by the indentation of the spheres into the soft membrane.

\section{Acknowledgments}

We thank Sid Nagel for insightful discussions and Helen Parks for performing initial tests of the gripping strength.   This work was supported by DARPA/DSO  through US Army Research Office grant W911NF-08-1-0140. Use of MRSEC shared experimental facilities at U of Chicago is gratefully acknowledged.

\section{Supplementary material}

In the main text we focused on vertical forces that pull along the central axis of the gripped object, i.e., along $\theta = 0$ in the sketch shown in Fig.~2a of the main text.  However, the object may also experience torques or off-axis forces, which might arise in situations when the object's center of mass lies off the central axis, when the gripper holding the object is rotated, or when the gripped object collides with an obstacle.  In this section we discuss the ability of the gripper to resist such off-axis forces and torques.

\begin{figure}                                                
\centerline{\includegraphics[width=3.15in]{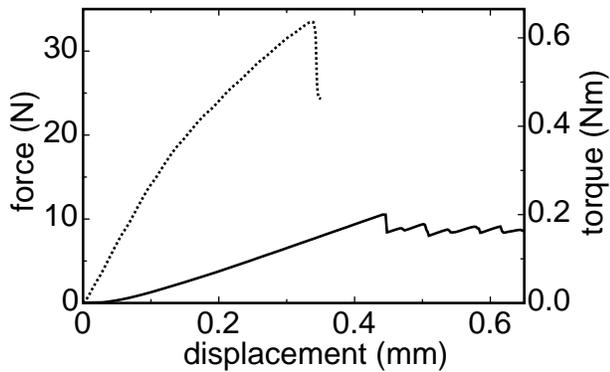}}
\caption{Force-displacement curves for off-axis forces and torques.  Dotted line:  Off-axis force applied to a solid sphere radially at $\theta=\pi/2$ with a gripper contact angle $\theta = 0.37\pi$ and $P_{jam}= 80$ kPa.  Solid line:  On-axis torque applied by rotating a key inserted into the sphere at $\theta=\pi$, a contact angle of $\theta=\pi/2$, and $P_{jam}= 54$ kPa.   The angular deflection has been converted to a displacement along the perimeter of the sphere by multiplying it by the sphere radius $R=19$ mm.  The torque scale on the right is the force scale multiplied by $R$, which corresponds to the maximum torque arm length of the friction applied by the gripper.   
}
\label{fig:stressstrainsixaxis}                                        
\end{figure}

\begin{figure}                                                
\centerline{\includegraphics[width=3.4in]{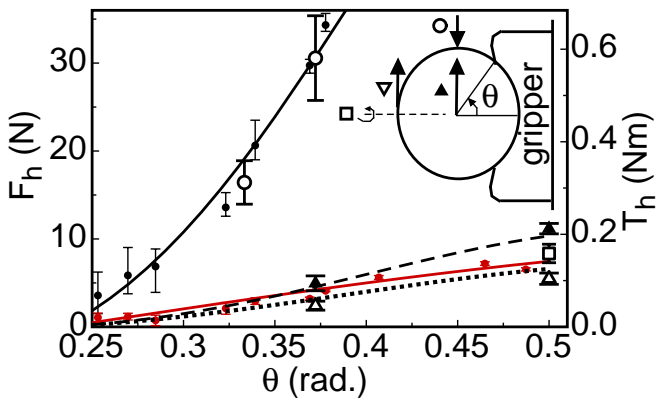}}
\caption{Maximum holding forces $F_h$ and torques $T_h$ on different axes as a function of contact angle $\theta$ between $\pi/4$ and $\pi/2$.  On-axis ($\theta=0$) forces on the center of mass are reproduced from Fig.~3 for a porous sphere (red solid circles)  and for a solid sphere (black solid circles), both at $P_{jam}=80$ kPa.    Solid lines give predictions as discussed in the main text.  Off-axis forces applied to a solid sphere radially at $\theta=\pi/2$  are shown as open black circles.  Torques are tested by pulling tangent to the edge of the sphere at radius $R$.  Solid triangles: on-axis torques, with force applied tangent to the surface at $\theta=\pi/2$ and $P_{jam}=80$ kPa.  Open triangles: off-axis torques, with force applied tangent to the surface at $\theta=\pi$ and $P_{jam}=80$ kPa.   Open square: on-axis torque, measured by a rotating a key inserted into the sphere at $\theta=\pi$.  Dashed and dotted lines:  model for on-axis and off-axis torques at $P_{jam}=80$ kPa, respectively, as discussed in the text.    Inset:  Diagram of the experimental setup, with arrows indicating the location and direction of applied forces for each measurement.  The torque scale on the right is the force scale multiplied by $R$.
}
\label{fig:sixaxis}                                        
\end{figure}

To measure the holding force of the gripper against off-axis forces we applied a force at $\theta = \pi/2$ in the radial direction toward the sphere center. To perform these tests, we embedded a solid test sphere of radius $R=19$ mm as usual by first pushing it into the gripper to a predetermined depth as shown in Fig.~2a. Next, the evacuated gripper holding the sphere was rotated by $\pi/2$ so the axis of cylindrical symmetry was horizontal (see inset to  Fig.~\ref{fig:sixaxis}). In this configuration,  force-displacement curves were taken with the Instron material tester by pushing a rod down onto the center of the sphere so it contacted at $\theta = \pi/2$.  An example force-displacement curve is shown in Fig.~\ref{fig:stressstrainsixaxis}.  The maximum force before failure was designated, as before, as the holding force, $F_h$.  Results for two contact angles are shown in Fig.~\ref{fig:sixaxis}.  It is seen that the holding force against these off-axis forces is just as strong as in the case as on-axis forces discussed in the main text.   

To test the holding force of the gripper against torques, we used the same sideways configuration for the gripper, and pulled on a string tied to the surface of the solid sphere, so the lever arm had the same length as the sphere radius. Data for the maximum holding torque, $T_h$, are shown in Fig.~\ref{fig:sixaxis} for on-axis torques applied at an angle $\pi/2$ relative to the central gripper axis, and off-axis torques applied at an angle $\pi$. While this method of applying torques also introduced some net force, that force is significantly weaker than the off- and on-axis holding forces. Therefore, it is safe to conclude that failure was due to the torque and not the net force. 

The forces for each torque measurement are in the range of the vertical holding force for frictional contact.  Analytically, the torque applied by the gripper to hold the target in place cannot come from suction since that only provides forces normal to the surface of the sphere. Instead, the full effect of friction can be applied perpendicular to the torque arm.  For on-axis torques, the torque arm around the center of the sphere is everywhere $R\sin\theta$, so the holding torque provided by friction is $T_h = F_f R\sin\theta$.  For off-axis torques, the torque arm varies along the O-ring as $R|\cos\phi|$  where $\phi$ is the azimuthal angle around the gripper axis.  The holding torque for off-axis torques is then  $T_h = F_f R \int_0^{2\pi} |(\cos\phi)| d\phi/(2\pi) = 2F_f R/\pi$.  For torques applied to the sphere at a radius of $R$, this predicts the same scale for the holding force $F_f$ as found for forces on the center of mass.  These models for the on- and off-axis torques are simultaneously fit to the data in Fig.~\ref{fig:sixaxis}, using  the scale factor $F_f$ as the only adjustable parameter (the resulting value of $F_f$ is about 50\% larger than the value found for the on-axis forces discussed in the main text;  however, the values of $F_f$ need not be the same because they are dependent on the contact area and this can vary as the gripper is deformed under different types of applied strain).  Generally, this model implies that the center of mass of the load can be located off the central axis ($\theta=0$) by as much as the mean radius of contact without a negative effect on the holding force.  

\begin{figure}                                                
\centerline{\includegraphics[width=2.5in]{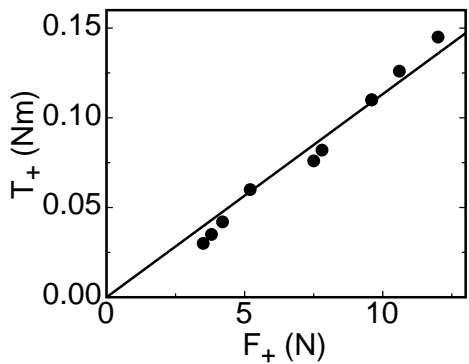}}
\caption{Contribution $T_+$ to the holding torque due to normal forces $F_+$ between the gripper and sphere of radius $R=19$ mm in addition to the holding torque provided by jamming.  Solid line:  a linear fit.   
}
\label{fig:normalforces}                                        
\end{figure}

In addition, we performed one more series of tests  in which the sphere was twisted around the central axis while an external normal force was applied.  Knowledge about such on-axis torque is relevant in cases where the gripper is pushed onto an object and then rotated, for example in order to open a door by twisting a round knob. In these situations, the normal force can provide additional friction and thus enhance the maximum sustainable torque.  For this measurement we used an Anton-Paar MCR 301 rheometer to measure both torque and angular deflection of a rotating tool while it applied a fixed normal force.   The measuring geometry was similar to Fig.~2a, except in this case the vertical rod was replaced with the rheometer tool which was rotated around its axis rather than pulled.  The rheometer tool was a cross-shaped key and was inserted into a matching slot on the top of a solid sphere with $R=19$ mm.  

To obtain a reference measurement with zero initial applied force in the normal direction, after pressing the sphere into the gripper and setting $P_{jam}$, we then back the key off slightly until the normal force reaches zero.  An example torque-deflection curve in this setup with zero normal force is shown in Fig.~\ref{fig:stressstrainsixaxis} for a contact angle of $\theta=\pi/2$ and $P_{jam}=54$ kPa.  The maximum resulting holding torque, $T_h$, is shown in Fig.~\ref{fig:sixaxis} (open square), where it is seen to be in good agreement with the other on-axis torque measurement when scaled by $P_{jam}$.  Several stick-slip events are apparent in the torque-deflection curve.  Note that, when the object is not also pulled out of the gripper, there is no permanent loss of grip after an initial slip.  In fact, the stress recovers with a slope similar to the initial stress buildup. 

To obtain results for different applied normal force values, we measured torque-deflection curves with the sphere pushed partially into the gripper by the rheometer.  Many of these tests correspond to contact angles below $\pi/4$.  Note that significant holding torques are found for small contact angles and even when $P_{jam}=0$.   We summarize these results by defining an extra holding torque, $T_+$, which is the excess over the torque obtained in an equivalent experiment with zero applied normal force.  This extra holding torque $T_+$ is plotted vs. the normal force at failure, $F_+$, in Fig.~\ref{fig:normalforces}.  We note that $F_+$ can differ from the applied normal force at the beginning of the measurement by a few Newtons due to deformations in the gripper -- it is specifically the normal force at failure that we find to be directly related to the holding torque. This additional contribution to the torque from the normal force is roughly proportional to  $F_+$, a relationship which holds for different values of contact angle and $P_{jam}$.  $T_+$ can be attributed to friction and expressed as $T_+ = F_+ R\sin\theta_+$. Here $\theta_+$ is a characteristic contact angle for the compression which is likely distributed over a wide area around the bottom of the sphere.  The proportionality between $T_+$ and $F_+$ indicated by the solid line in Fig.~\ref{fig:normalforces} corresponds to $\theta_+ = 0.4\pi$.  These results show that, given enough normal force, the gripper can achieve holding torques well above those produced by jamming alone.

Because we argued that the frictional force is more than enough to lift any object on the scale of a few cm or less, our holding torque model suggests that the grip will not fail when the lifted object is rotated or picked up off-center.  Furthermore, the displacement of the target in the gripper under off-axis forces shown in  Fig.~\ref{fig:stressstrainsixaxis} was only 7 $\mu$m at the point where the holding force equals the weight.  Similarly, the stiffness to torque in the rheometer measurements was so large that even if the weight was distributed at the edge of the sphere, tilting would cause a deflection of only 0.001 radians or an edge displacement of 20 $\mu$m.   These results show that the gripper is capable of precision holding during changes in orientation.  

This discussion focusing on spheres can be extended to other shapes.  The extreme case where there is a long lever arm on which torques can be most easily applied is a horizontal rod longer than the gripper.  The above analysis implies that the center of mass of the rod only needs to be somewhere within the space directly below the gripper for optimal gripping performance.  If such a rod were gripped and then pushed along its axis, we would expect the gripper to respond with friction only since in this case suction would be entirely perpendicular to the applied force.  However, suction could be operable in holding against torques in directions perpendicular to the rod axis, in which the lever arm is largest.  For other shapes we would expect friction will provide a holding force against forces on any axis and some component of suction may be able to provide some holding force depending on target shape and force orientation.

\end{document}